\documentclass[11pt]{article}
\newcommand{\be}{\begin{equation}}
\newcommand{\ee}{\end{equation}}
\newcommand{\bea}{\begin{eqnarray}}
\newcommand{\eea}{\end{eqnarray}}

\newcommand{\sech}{{\rm sech}}

\begin{document}

\vspace{0.5in}
\begin{center}
{\LARGE{\bf 
Superposed Hyperbolic Kink and Pulse Solutions of Coupled $\phi^4$, NLS and MKdV Equations}}
\end{center}

\begin{center}
{\LARGE{\bf Avinash Khare}} \\
{Physics Department, Savitribai Phule Pune University \\
 Pune 411007, India}
\end{center}

\begin{center}
{\LARGE{\bf Avadh Saxena}} \\ 
{Theoretical Division and Center for Nonlinear Studies, 
Los Alamos National Laboratory, Los Alamos, New Mexico 87545, USA}
\end{center}

\vspace{0.9in}
\noindent{\bf {Abstract:}}

We obtain novel solutions of a coupled $\phi^4$, a coupled nonlinear 
Schr\"odinger (NLS) and a coupled modified Korteweg de Vries (MKdV) 
model which can be re-expressed as a linear superposition of either the 
sum or the difference of two hyperbolic kink or two hyperbolic pulse solutions. 
These results demonstrate that the notion of superposed solutions extends to 
coupled nonlinear equations as well. 

\section{Introduction}
 
Superposition principle is the hallmark of linear theories and it has 
helped us in understanding various features of such theories. In contrast, 
because of the nonlinear term, the nonlinear theories lack such a 
superposition principle. This we believe is one of the reasons 
why until now we only have a limited knowledge of such theories. 
Undoubtedly, during the last five decades substantial progress has been 
made in understanding several novel features of a number of nonlinear models 
but even at present new features of nonlinear theories are being uncovered. 
In a recent paper \cite{ks22} we have  shown that a large number of 
nonlinear equations, including both symmetric and asymmetric $\phi^4$, 
MKdV, NLS and several other nonlinear equations, many of which
have found wide application in physics, admit superposed periodic  
kink and periodic pulse solutions. 

By {\it superposed} here we mean that a 
solution is a linear combination of two kink solutions or two pulse 
solutions. However, only one hyperbolic superposed solution was obtained
by us in some of these models as the difference of two kink solutions. 
In fact, even before us a number of previous researchers \cite{tan, sma, 
dashen, campbell, saxena, thies} had obtained such a superposed kink 
solution in a variety of physical models. However, to the best of our 
knowledge, to date no one has been able to obtain solutions which can 
be re-expressed either as the sum of two (hyperbolic) kink solutions 
or the sum or difference of two pulse solutions. The purpose of this paper 
is to partially fill this gap. 

We present three well known {\it coupled} models, i.e. a coupled $\phi^4$, a 
coupled NLS and a coupled MKdV, which have found widespread applications 
in ferroelectric \cite{abel} and multiferroic \cite{curnoe} phase transiitons,  
signal propagation in optical fibers \cite{mecozzi}, etc.  We then obtain novel 
solutions which can be re-expressed either as the 
sum or the difference of two (hyperbolic) kink or two pulse solutions. Needless 
to say that these models also admit novel solutions which can be re-expressed
as the sum or the difference of two periodic kink or two periodic pulse solutions. 
However, in this paper we only confine ourselves to the superposition of  hyperbolic 
kink and pulse solutions. In a subsequent paper we hope to address the question 
of the superposition of periodic kink and pulse solutions in these models.

The plan of the paper is the following. In Sec. II, we discuss a coupled $\phi^4$ 
model which has received attention previously \cite{coupledphi4} and
obtain seven  solutions which can be re-expressed as either the 
sum or the difference of two kink or two pulse solutions. 
In Sec. III, we discuss a coupled NLS model, also popularly known
as the Manakov model \cite{manakov} and show that it admits seven solutions
(which are similar to the corresponding solutions of the coupled $\phi^4$ model) 
which can be re-expressed as the sum or the difference of two kink or two pulse 
solutions. In Sec. IV we discuss a coupled MKdV model \cite{coupledmkdv} 
and show that this model admits four solutions which can be re-expressed as
the sum or the difference of  two kink solutions. Finally, in Sec. V we summarize  
our main results and point out some of the open problems.

\section{Novel Superposed Solutions of a Coupled $\phi^4$ Model}

Let us consider the following coupled $\phi^4$ equations \cite{coupledphi4}  
\be\label{1}
\phi_{1xx} = a_1 \phi_1 +[b_1 \phi_{1}^{2}+ d_1 \phi_{2}^2]\phi_1\,,
\ee
\be\label{2} 
\phi_{2xx} = a_2 \phi_2 +[b_2 \phi_{1}^{2}+ d_2 \phi_{2}^2]\phi_2\,.
\ee
Before we discuss the superposed kink and pulse solutions, let us note
that the coupled Eqs. (\ref{1}) and (\ref{2}) admit the following kink
and pulse solutions.

It is well known that Eqs. (\ref{1}) and (\ref{2}) admit a kink 
solution in both $\phi_1$ and $\phi_2$, i.e.
\be\label{3}
\phi_1(x) = A_1 \tanh(\beta x)\,,~~\phi_2 = A_2 \tanh(\beta x)\,,
\ee
provided
\be\label{4}
a_1 = a_2 = -2\beta^2\,,~~b_1 A_{1}^{2} + d_1 A_{2}^{2} = 
b_2 A_{1}^{2} + d_{2} A_{2}^{2} = 2\beta^2\,.
\ee

It is also known that Eqs. (\ref{1}) and (\ref{2}) admit a pulse 
solution in both $\phi_1$ and $\phi_2$, i.e.
\be\label{5}
\phi_1(x) = A_1 \sech(\beta x)\,,~~\phi_2 = A_2 \sech(\beta x)\,,
\ee
provided
\be\label{6}
a_1 = a_2 = \beta^2\,,~~b_1 A_{1}^{2} + d_1 A_{2}^{2} = 
b_2 A_{1}^{2} + d_{2} A_{2}^{2} = -2\beta^2\,.
\ee

Finally, Eqs. (\ref{1}) and (\ref{2}) are also known to admit a mixed
kink-pulse solution, i.e. say a kink solution in $\phi_1$ and a
pulse solution in $\phi_2$ or vice versa. In particular, they
admit
\be\label{7}
\phi_1(x) = A_1 \tanh(\beta x)\,,~~\phi_2 = A_2 \sech(\beta x)\,,
\ee
provided
\bea\label{8}
&&b_1 A_{1}^{2} - d_1 A_{2}^{2} = b_2 A_{1}^{2} 
-d_{2} A_{2}^{2} = 2\beta^2\,, \nonumber \\ 
&&b_2 A_{1}^2 = \beta^2 - a_2\,,~~d_1 A_{2}^{2} 
= -a_1 - 2 \beta^2\,.
\eea

We now show that the coupled Eqs. (\ref{1}) and (\ref{2}) 
in fact also admit novel solutions which can be re-expressed
as the sum or the difference of the above kink and pulse solutions.

Let us first note that in case $\phi_2(x) = \alpha\phi_1(x)$, 
where $\alpha$ is a real number, then  
Eq. (\ref{2}) is identical to Eq. (\ref{1}) provided
\be\label{9}
a_2 = a_1\,,~~b_1 +\alpha^2 d_1 = b_2 +\alpha^2 d_2\,,
\ee
and in this case we only need to solve Eq. (\ref{1}) which 
as we have recently shown \cite{ks22} does not admit superposed
hyperbolic solutions. 

We now show that instead when $\phi_2$ and $\phi_1$ are not 
proportional to each other then the coupled Eqs. (\ref{1}) and
(\ref{2}) admit seven distinct solutions which can be re-expressed
as the sum or the difference of either two kink or two pulse 
solutions.

{\bf Superposed Solution I}

It is easy to check that the coupled Eqs. (\ref{1}) and 
(\ref{2}) admit the superposed hyperbolic solution

\be\label{10}
\phi_1(x) = \frac{A}{B+\cosh^2(\beta x)}\,,
~~\phi_2(x) = \frac{D \cosh(\beta x) \sinh(\beta x)}
{B+\cosh^2(\beta x)}\,,~~B > 0\,,
\ee
provided
\bea\label{11}
&&a_1 = a_2 = 2\beta^2\,,~~b_1 A^2 = -2B(B+1)\beta^2\,,~~d_1 D^2 
= -6\beta^2\,, \nonumber \\
&&b_2 A^2 = -6B(B+1) \beta^2\,,~~d_2 D^2 = -2 \beta^2\,.
\eea
Thus for this solution $b_1,d_1, b_2,d_2$ are all negative while $a_1,a_2$
are positive.

Now on making use of the addition theorem for $\tanh(a+b)$, 
one can derive two novel identities
\be\label{12}
\tanh(x+\Delta)-\tanh(x-\Delta) = 
\frac{\sinh(2\Delta)}{B+\cosh^2(x)}\,,~~B = \sinh^2(\Delta)\,,
\ee
\be\label{13}
\tanh(x+\Delta)+\tanh(x-\Delta) = 
\frac{2\sinh(x)\cosh(x)}{B+\cosh^2(x)}\,,~~B = \sinh^2(\Delta)\,.
\ee
On comparing the solution (\ref{10}) with the identities (\ref{12}) 
and (\ref{13}), solution I as given by Eq. (\ref{10}) can be
re-expressed as
\bea\label{14}
&&\phi_1(x) = \frac{\beta}{\sqrt{2|b_1|}}[\tanh(\beta x + \Delta)
-\tanh(\beta x - \Delta)]\,, \nonumber \\
&&\phi_2(x) = \frac{\sqrt{3}\beta}{\sqrt{2|d_1|}}
[\tanh(\beta x + \Delta)+\tanh(\beta x - \Delta)]\,,
\eea
where $\sinh^2(\Delta) = B$.

{\bf Superposed Solution II}

It is easy to check that the coupled Eqs. (\ref{1}) and 
(\ref{2}) admit the superposed hyperbolic solution
\be\label{15}
\phi_1(x) = \frac{A}{B+\cosh^2(\beta x)}\,,
~~\phi_2(x) = \frac{D \sinh(\beta x)}
{B+\cosh^2(\beta x)}\,,~~B > 0\,,
\ee
provided
\bea\label{16}
&&a_1 = -4\beta^2\,,~~b_1 A^2 = 2(2B^2+5B+3)\beta^2\,,~~d_1 D^2 
= 6(2B+1)\beta^2\,, \nonumber \\
&&a_2 = -\beta^2\,,~~b_2 A^2 = 6(B+1) \beta^2\,,~~d_2 \beta^2 
= 2(3+4B)\beta^2\,.
\eea
Thus for this solution $b_1,d_1, b_2,d_2$ are all positive while $a_1,a_2$
are negative.

Now on making use of the addition theorem for $\sech(a+b)$, one can derive two
novel identities
\be\label{17}
\sech(x-\Delta)-\sech(x+\Delta) = \frac{2\sinh(x)\sinh(\Delta)}{B+\cosh^2(x)}\,,
~~B = \sinh^2(\Delta)\,,
\ee
\be\label{18}
\sech(x+\Delta)+\sech(x-\Delta) = \frac{2\cosh(\Delta)\cosh(x)}{B+\cosh^2(x)}\,,
~~B = \sinh^2(\Delta)\,.
\ee
On comparing the solution (\ref{15}) with the identities (\ref{12}) 
and (\ref{17}), solution II as given by Eq. (\ref{15}) can be
re-expressed as
\bea\label{19}
&&\phi_1(x) = \frac{\sqrt{3}\beta}{\sqrt{2b_2} \sinh(\Delta)}
[\tanh(\beta x + \Delta) -\tanh(\beta x - \Delta)]\,, \nonumber \\
&&\phi_2(x) = \frac{\sqrt{3 \cosh(2\Delta)}\beta}
{\sqrt{2 d_1}\sinh(\Delta)}
[\sech(\beta x - \Delta)- \sech(\beta x +\Delta)]\,,
\eea
where $B = \sinh^2(\Delta)$.

{\bf Superposed Solution III}

It is easy to check that the coupled Eqs. (\ref{1}) and 
(\ref{2}) admit the superposed hyperbolic solution
\be\label{20}
\phi_1(x) = \frac{A}{B+\cosh^2(\beta x)}\,,
~~\phi_2(x) = \frac{D \cosh(\beta x)}
{B+\cosh^2(\beta x)}\,,~~B > 0\,,
\ee
provided
\bea\label{21}
&&a_1 = -4\beta^2\,,~~b_1 A^2 = 2B(2B-1)\beta^2\,,~~d_1 D^2 
= 6(2B+1)\beta^2\,, \nonumber \\
&&a_2 = -\beta^2\,,~~b_2 A^2 = -6 B \beta^2\,,~~d_2 \beta^2 
= 2(1+4B)\beta^2\,.
\eea
Thus for this solution while $d_1,d_2 < 0$, $a_1, a_2, b_2 < 0$ while
$b_1 > $ $(< )$ 0 depending on if $B >$ $ (<)$ 1/2 while $b_1 = 0$ at 
$B = 1/2$. 

On comparing the solution (\ref{21}) with the identities (\ref{12}) 
and (\ref{18}), solution III as given by Eq. (\ref{20}) can be
re-expressed as
\bea\label{22}
&&\phi_1(x) = \frac{\sqrt{3}\beta}{\sqrt{2|b_2|} \cosh(\Delta)}
[\tanh(\beta x + \Delta) -\tanh(\beta x - \Delta)]\,, \nonumber \\
&&\phi_2(x) = \frac{\sqrt{3 \cosh(2\Delta)}\beta}{\sqrt{2 d_1}\cosh(\Delta)}
[\sech(\beta x + \Delta) + \sech(\beta x - \Delta)]\,,
\eea
where $B = \sinh^2(\Delta)$.

{\bf Superposed Solution IV}

It is easy to check that the
coupled Eqs. (\ref{1}) and (\ref{2}) admit the superposed 
hyperbolic solution
\be\label{23}
\phi_1(x) = \frac{A\sinh(\beta x)}{B+\cosh^2(\beta x)}\,,
~~\phi_2(x) = \frac{D \sinh(\beta x) \cosh(\beta x)}
{B+\cosh^2(\beta x)}\,,~~B > 0\,,
\ee
provided
\bea\label{24}
&&a_1 = \frac{(5-B)\beta^2}{(1+B)}\,,~~b_1 A^2 = 2(2B^2+5B+3)\beta^2\,,
~~d_1 D^2 = 6(2B+1)\beta^2\,, \nonumber \\
&&a_2 = -\beta^2\,,~~b_2 A^2 = 6(B+1) \beta^2\,,~~d_2 \beta^2 
= 2(3+4B)\beta^2\,.
\eea
Thus for this solution $b_1,d_1, b_2,d_2$ are all positive while $a_1,a_2$
are negative.

On making use of the novel identities (\ref{13}) and (\ref{17}), one
can then re-express solution IV as given by Eq. (\ref{23}) as
\bea\label{25}
&&\phi_1(x) = \frac{\sqrt{3}\beta}{\sqrt{2b_2} \coth(\Delta)}
[\sech(\beta x - \Delta) -\sech(\beta x + \Delta)]\,, \nonumber \\
&&\phi_2(x) = \frac{\sqrt{3 \cosh(2\Delta)}\beta}
{\sqrt{2 d_1}}
[\tanh(\beta x + \Delta) + \tanh(\beta x -\Delta)]\,,
\eea
where $B = \sinh^2(\Delta)$.

{\bf Superposed Solution V}

It is easy to check that the coupled Eqs. (\ref{1}) and 
(\ref{2}) admit the superposed hyperbolic solution
\be\label{26}
\phi_1(x) = \frac{A\cosh(\beta x)}{B+\cosh^2(\beta x)}\,,
~~\phi_2(x) = \frac{D \sinh(\beta x) \cosh(\beta x)}
{B+\cosh^2(\beta x)}\,,~~B > 0\,,
\ee
provided
\bea\label{27}
&&a_1 = -\frac{(6+B)\beta^2}{B}\,,~~b_1 A^2 = 
\frac{2(B+1)(4B+3)\beta^2}{B}\,, \nonumber \\
&&d_1 D^2 = \frac{6 \beta^2}{B}\,,
~~a_2 = -\frac{2(2B+3)beta^2}{B}\,, \nonumber \\
&&b_2 A^2 = \frac{2(2B^2+13B+3)\beta^2}{B}\,,~~d_2  
= 2(3+4B)\beta^2\,.
\eea
Thus for this solution $b_1,d_1, b_2,d_2$ are all positive while $a_1,a_2$
are negative.

On making use of the novel identities (\ref{13}) and (\ref{18}), one
can then re-express solution V as given by Eq. (\ref{26}) 
re-expressed as
\bea\label{28}
&&\phi_1(x) = \frac{\sqrt{4\sinh^2(\Delta)+3}\beta}
{\sqrt{2b_1} \sinh(\Delta)}
[\sech(\beta x + \Delta) +\sech(\beta x - \Delta)]\,, \nonumber \\
&&\phi_2(x) = \frac{\sqrt{4\sinh^2(\Delta)+3}\beta}
{\sqrt{2 d_2}\sinh(\Delta)}
[\tanh(\beta x + \Delta)+ \tanh(\beta x -\Delta)]\,,
\eea
where $B = \sinh^2(\Delta)$.

{\bf Superposed Solution VI}

It is easy to check that the coupled Eqs. (\ref{1}) and 
(\ref{2}) admit the superposed hyperbolic solution
\be\label{29}
\phi_1(x) = \frac{A\sinh(\beta x)}{B+\cosh^2(\beta x)}\,,
~~\phi_2(x) = \frac{D \cosh(\beta x)}
{B+\cosh^2(\beta x)}\,,~~B > 0\,,
\ee
provided
\bea\label{30}
&&a_1 = -\beta^2\,,~~b_1 A^2 = 2B \beta^2\,,~~d_1 D^2 
= 6(B+1)\beta^2\,, \nonumber \\
&&a_2 = -\beta^2\,,~~b_2 A^2 = 6 B \beta^2\,,~~d_2 \beta^2 
= 2(B+1)\beta^2\,.
\eea
Thus for this solution $b_1,d_1, b_2,d_2$ are all positive while $a_1,a_2$
are negative.

On making use of the novel identities (\ref{17}) and (\ref{18}), one
can then re-express solution VI as given by Eq. (\ref{29}) as
\bea\label{31}
&&\phi_1(x) = \frac{\beta}{\sqrt{2 b_1}}
[\sech(\beta x - \Delta) -\sech(\beta x + \Delta)]\,, \nonumber \\
&&\phi_2(x) = \frac{\beta}{\sqrt{2 d_2}}
[\sech(\beta x + \Delta)+ \sech(\beta x -\Delta)]\,, 
\eea
where $B = \sinh^2(\Delta)$. 

{\bf Superposed Solution VII}

It is easy to check that the coupled Eqs. (\ref{1}) and 
(\ref{2}) admit the superposed hyperbolic solution

\be\label{32}
\phi_1(x) = 1- \frac{A}{B+\cosh^2(\beta x)}\,,
~~\phi_2(x) = \frac{D}{B+\cosh^2(\beta x)}\,,~~A, B, D > 0\,,
\ee
provided
\bea\label{33}
&&b_1 = -a_1 = 2\beta^2\,,~~A = \frac{3(1+2B)+\sqrt{8+(2B+1)^2}}{4}\,,
\nonumber \\
&&d_1 D^2 = [\sqrt{8+(2B+1)^2}-(2B+1)] \frac{3A\beta^2}{2}\,,
~~b_2 + a_2 = 4\beta^2\,, \nonumber \\
&&b_2 A = 3(2B+1) \beta^2\,,
~~b_2 A^2 + d_2 D^2 = 8B(B+1) \beta^2\,.
\eea
Thus for this solution $b_1,d_1, a_2, b_2$ are all positive, 
$a_1 < 0$ while the sign of $d_2$ depends on the value of $B$.  

On comparing the solution (\ref{32}) with the identity (\ref{12}), 
solution VII as given by Eq. (\ref{32}) can be
re-expressed as
\bea\label{34}
&&\phi_1(x) = 1- \frac{\beta}{\sqrt{2|b_1|}}[\tanh(\beta x + \Delta)
-\tanh(\beta x - \Delta)]\,, \nonumber \\
&&\phi_2(x) = \frac{\sqrt{3}\beta}{2\sqrt{d_1}} K
[\tanh(\beta x + \Delta)-\tanh(\beta x - \Delta)]\,,
\eea
where $\sinh^2(\Delta) = B$ and 
\be\label{35}
K = \bigg [\cosh(2\Delta)\sqrt{8+\cosh^2(2\Delta)}
-[2+\cosh^2(2\Delta)] \bigg ]^{1/2}\,.
\ee

\section{Hyperbolic Superposed Solutions of the Coupled NLS Model}

Let us consider the following coupled NLS equations \cite{manakov} 
\be\label{2.1}
i u_{1t} + u_{1xx} +[g_{11} |u_1|^2 +g_{12} |u_2|^2] u_1\,,
\ee
\be\label{2.2}
i u_{2t} + u_{2xx} +[g_{21} |u_1|^2 +g_{22} |u_2|^2] u_2\,. 
\ee

Before we discuss the superposed kink and pulse solutions, let us note
that the coupled Eqs. (\ref{2.1}) and (\ref{2.2}) admit the following kink
and pulse solutions. In particular, it is well known that Eqs. 
(\ref{2.1}) and (\ref{2.2}) admit a kink 
solution in both $u_1$ and $u_2$, i.e.
\be\label{2.3}
u_1(x,t) = A_1 e^{i\omega_1 t}\tanh(\beta x)\,,
~~u_2(x,t) = A_2 e^{i\omega t} \tanh(\beta x)\,,
\ee
provided
\be\label{2.4}
\omega_1 = \omega_2 = -2\beta^2\,,~~g_{11} A_{1}^{2} 
+ g_{12} A_{2}^{2} = g_{21} A_{1}^{2} 
+ g_{22} A_{2}^{2} = 2\beta^2\,.
\ee

It is also known that Eqs. (\ref{2.1}) and (\ref{2.2}) admit a pulse 
solution in both $u_1$ and $u_2$, i.e.
\be\label{2.5}
u_1(x,t) = A_1 e^{i\omega_1 t} \sech(\beta x)\,,
~~u_2(x,t) = A_2 e^{i\omega_2 t} \sech(\beta x)\,,
\ee
provided
\be\label{2.6}
\omega_1 = \omega_2 = \beta^2\,,~~g_{11} A_{1}^{2} 
+ g_{12} A_{2}^{2} = g_{21} A_{1}^{2} 
+ g_{22} A_{2}^{2} = -2\beta^2\,.
\ee

Finally, Eqs. (\ref{2.1}) and (\ref{2.2}) are also known to admit 
a  mixed kink-pulse solution, i.e. say a kink solution in 
$u_1$ and a pulse solution in $u_2$ or vice versa. In 
particular, they admit
\be\label{2.7}
u_1(x,t) = A_1 e^{i\omega_1 t} \tanh(\beta x)\,,
~~u_2(x,t) = A_2 e^{i\omega_2 t} \sech(\beta x)\,,
\ee
provided
\bea\label{2.8}
&&g_{11} A_{1}^{2} - g_{12} A_{2}^{2} 
= g_{21} A_{1}^{2} -g_{22} A_{2}^{2} = 2\beta^2\,, \nonumber \\ 
&&g_{21} A_{1}^2 = \beta^2 - \omega_2\,,~~g_{12} A_{2}^{2} 
= -\omega_1 - 2 \beta^2\,.
\eea

We now show that the coupled Eqs. (\ref{2.1}) and (\ref{2.2}) 
in fact also admit novel solutions which can be re-expressed
as the sum or the difference of the above kink and pulse solutions.

Let us first note that in case $u_2(x,t) = \alpha u_1(x,t)$, 
where $\alpha$ is a real number, then  
Eq. (\ref{2.2}) is identical to Eq. (\ref{2.1}) provided
\be\label{2.9}
g_{11} +\alpha^2 g_{12} = g_{21} +\alpha^2 g_{22}\,,
\ee
and in this case we only need to solve Eq. (\ref{2.1}) which 
as we have recently shown \cite{ks22} does not admit superposed
hyperbolic solutions. 

We now show that instead when $u_2$ and $u_1$ are not 
proportional to each other then the coupled Eqs. (\ref{2.1}) and
(\ref{2.2}) admit seven distinct solutions which can be re-expressed
as the sum or the difference of either two kink or two pulse solutions.

{\bf Superposed Solution I}

It is easy to check that the coupled Eqs. (\ref{2.1}) and 
(\ref{2.2}) admit the superposed hyperbolic solution

\be\label{2.10}
u_1(x,t) = e^{i\omega_1 t} \frac{A}{B+\cosh^2(\beta x)}\,,
~~u_2(x,t) = e^{i\omega_2 t} \frac{D \cosh(\beta x) \sinh(\beta x)}
{B+\cosh^2(\beta x)}\,,~~B > 0\,,
\ee
provided
\bea\label{2.11}
&&\omega_1 = \omega_2 = 2\beta^2\,,~~g_{11} A^2 
= -2B(B+1)\beta^2\,,~~g_{12} D^2 = -6\beta^2\,, \nonumber \\
&&g_{21} A^2 = -6B(B+1) \beta^2\,,
~~g_{22} D^2 = -2 \beta^2\,.
\eea
Thus for this solution $g_{11},g_{12}, g_{21}, g_{22}$ 
are all negative while $\omega_1 = \omega_2$
are positive.

On comparing the solution (\ref{2.10}) with the identities (\ref{12}) 
and (\ref{13}), the solution I as given by Eq. (\ref{2.10}) can be
re-expressed as
\bea\label{2.14}
&&u_1(x,t) = e^{i\omega_1 t} \frac{\beta}{\sqrt{2|g_{11}|}}
[\tanh(\beta x + \Delta) -\tanh(\beta x - \Delta)]\,, 
\nonumber \\
&&u_2(x,t) = e^{i\omega_2 t} \frac{\sqrt{3}\beta}{\sqrt{2|g_{12}|}}
[\tanh(\beta x + \Delta)+\tanh(\beta x - \Delta)]\,,
\eea
where $\sinh^2(\Delta) = B$.

{\bf Superposed Solution II}

It is easy to check that the coupled Eqs. (\ref{2.1}) and 
(\ref{2.2}) admit the superposed hyperbolic solution
\be\label{2.15}
u_1(x,t) = e^{i\omega t} \frac{A}{B+\cosh^2(\beta x)}\,,
~~u_2(x,t) = e^{i\omega_2 t} \frac{D \sinh(\beta x)}
{B+\cosh^2(\beta x)}\,,~~B > 0\,,
\ee
provided
\bea\label{2.16}
&&\omega_1 = -4\beta^2\,,~~g_{11} A^2 = 2(2B^2+5B+3)\beta^2\,,
~~g_{12} D^2 = 6(2B+1)\beta^2\,, \nonumber \\
&&\omega_2 = -\beta^2\,,~~g_{21} A^2 = 6(B+1) \beta^2\,,
~~g_{22} \beta^2 = 2(3+4B)\beta^2\,.
\eea
Thus for this solution $g_{11},g_{12}, g_{21}, g_{22}$ 
are all positive while $\omega_1, \omega_2$ are negative.

On comparing the solution (\ref{2.15}) with the identities (\ref{12}) 
and (\ref{17}), solution II as given by Eq. (\ref{2.15}) can be
re-expressed as
\bea\label{2.19}
&&u_1(x,t) = e^{i\omega_1 t} \frac{\sqrt{3}\beta}
{\sqrt{2g_{21}} \sinh(\Delta)}
[\tanh(\beta x + \Delta) -\tanh(\beta x - \Delta)]\,, \nonumber \\
&&u_2(x,t) = e^{i\omega_2 t} \frac{\sqrt{3 \cosh(2\Delta)}\beta}
{\sqrt{2 g_{12}}\sinh(\Delta)}
[\sech(\beta x - \Delta)- \sech(\beta x +\Delta)]\,,
\eea
where $B = \sinh^2(\Delta)$.

{\bf Superposed Solution III}

It is easy to check that the coupled Eqs. (\ref{2.1}) and 
(\ref{2.2}) admit the superposed hyperbolic solution
\be\label{2.20}
u_1(x,t) = e^{i\omega_1 t} \frac{A}{B+\cosh^2(\beta x)}\,,
~~u_2(x,t) = e^{i\omega_2 t} \frac{D \cosh(\beta x)}
{B+\cosh^2(\beta x)}\,,~~B > 0\,,
\ee
provided
\bea\label{2.21}
&&\omega_1 = -4\beta^2\,,~~g_{11} A^2 = 2B(2B-1)\beta^2\,,
~~g_{12} D^2 = 6(2B+1)\beta^2\,, \nonumber \\
&&\omega_2 = -\beta^2\,,~~g_{21} A^2 = -6 B \beta^2\,,
~~g_{22} \beta^2 = 2(1+4B)\beta^2\,.
\eea
Thus for this solution while $g_{12},g_{22} < 0$, 
$\omega_1, \omega_2, g_{21} < 0$ while
$g_{11} >$ $(< )$ 0 depending on if $B >$ $(<)$ 1/2 
while $g_{11} = 0$ at $B = 1/2$. 

On comparing the solution (\ref{2.21}) with the identities (\ref{12}) 
and (\ref{18}), solution III as given by Eq. (\ref{2.20}) can be
re-expressed as
\bea\label{2.22}
&&u_1(x,t) = e^{i\omega_1 t} \frac{\sqrt{3}\beta}
{\sqrt{2|g_{21}|} \cosh(\Delta)}
[\tanh(\beta x + \Delta) -\tanh(\beta x - \Delta)]\,, 
\nonumber \\
&&u_2(x,t) = e^{i\omega_2 t} 
\frac{\sqrt{3 \cosh(2\Delta)}\beta}{\sqrt{2 g_{12}}\cosh(\Delta)}
[\sech(\beta x + \Delta) + \sech(\beta x - \Delta)]\,,
\eea
where $B = \sinh^2(\Delta)$.

{\bf Superposed Solution IV}

It is easy to check that the
coupled Eqs. (\ref{2.1}) and (\ref{2.2}) admit the superposed 
hyperbolic solution
\be\label{2.23}
u_1(x,t) = e^{i\omega_1 t} \frac{A\sinh(\beta x)}
{B+\cosh^2(\beta x)}\,,
~~u_2(x,t) = e^{i\omega_2 t} \frac{D \sinh(\beta x) \cosh(\beta x)}
{B+\cosh^2(\beta x)}\,,~~B > 0\,,
\ee
provided
\bea\label{2.24}
&&\omega_1 = \frac{(5-B)\beta^2}{(1+B)}\,,~~g_{11} A^2 
= 2(2B^2+5B+3)\beta^2\,,~~g_{12} D^2 = 6(2B+1)\beta^2\,, 
\nonumber \\
&&\omega_2 = -\beta^2\,,~~g_{21} A^2 = 6(B+1) \beta^2\,,
~~g_{22} \beta^2 = 2(3+4B)\beta^2\,.
\eea
Thus for this solution $g_{11}, g_{21}, g_{21}, g_{22}$ 
are all positive while $\omega_1, \omega_2$ are negative.

On making use of the novel identities (\ref{13}) and (\ref{17}), one
can then re-express solution IV as given by Eq. (\ref{2.23}) as
\bea\label{2.25}
&&u_1(x,t) = e^{i\omega_1 t} \frac{\sqrt{3}\beta}
{\sqrt{2 g_{21}} \coth(\Delta)}
[\sech(\beta x - \Delta) -\sech(\beta x + \Delta)]\,, \nonumber \\
&&u_2(x,t) = e^{i\omega_2 t} \frac{\sqrt{3 \cosh(2\Delta)}\beta}
{\sqrt{2 g_{12}}} [\tanh(\beta x + \Delta) + \tanh(\beta x -\Delta)]\,,
\eea
where $B = \sinh^2(\Delta)$.
\vskip 0.4cm 

{\bf Superposed Solution V}

It is easy to check that the coupled Eqs. (\ref{2.1}) and 
(\ref{2.2}) admit the superposed hyperbolic solution
\be\label{2.26}
u_1(x,t) = e^{i\omega_1 t} \frac{A\cosh(\beta x)}
{B+\cosh^2(\beta x)}\,,~~u_2(x,t) = e^{i\omega_2 t} 
\frac{D \sinh(\beta x) \cosh(\beta x)}{B+\cosh^2(\beta x)}\,,~~B > 0\,,
\ee
provided
\bea\label{2.27}
&&\omega_1 = -\frac{(6+B)\beta^2}{B}\,,~~g_{11} A^2 = 
\frac{2(B+1)(4B+3)\beta^2}{B}\,, \nonumber \\
&&g_{12} D^2 = \frac{6 \beta^2}{B}\,,~~
\omega_2 = -\frac{2(2B+3)beta^2}{B}\,, \nonumber \\
&&g_{21} A^2 = \frac{2(2B^2+13B+3)\beta^2}{B}\,,
~~g_{22} = 2(3+4B)\beta^2\,.
\eea
Thus for this solution $g_{11}, g_{12}, g_{21}, g_{22}$ 
are all positive while $\omega_1, \omega_2$ are negative.

On making use of the novel identities (\ref{13}) and (\ref{18}), one
can then re-express solution V as given by Eq. (\ref{2.26}) as
\bea\label{2.28}
&&u_1(x,t) = \frac{\sqrt{4\sinh^2(\Delta)+3}\beta}
{\sqrt{2 g_{11}} \sinh(\Delta)}
[\sech(\beta x + \Delta) +\sech(\beta x - \Delta)]\,, 
\nonumber \\
&&u_2(x,t) = \frac{\sqrt{4\sinh^2(\Delta)+3}\beta}
{\sqrt{2 g_{22}}\sinh(\Delta)}
[\tanh(\beta x + \Delta)+ \tanh(\beta x -\Delta)]\,,
\eea
where $B = \sinh^2(\Delta)$.

{\bf Superposed Solution VI}

It is easy to check that the coupled Eqs. (\ref{2.1}) and 
(\ref{2.2}) admit the superposed hyperbolic solution
\be\label{2.29}
u_1(x,t) = e^{i\omega_1 t} \frac{A\sinh(\beta x)}
{B+\cosh^2(\beta x)}\,,~~u_2(x,t) = e^{i\omega_2 t} 
\frac{D \cosh(\beta x)}{B+\cosh^2(\beta x)}\,,~~B > 0\,,
\ee
provided
\bea\label{2.30}
&&\omega_1 = -\beta^2\,,~~g_{11} A^2 = 2B \beta^2\,,
~~g_{12} D^2 = 6(B+1)\beta^2\,, \nonumber \\
&&\omega_2 = -\beta^2\,,~~g_{21} A^2 = 6 B \beta^2\,,
~~g_{22} \beta^2 = 2(B+1)\beta^2\,.
\eea
Thus for this solution $g_{11},g_{12}, g_{21}, g_{22} $ 
are all positive while $\omega_1, \omega_2$ are negative.

On making use of the novel identities (\ref{17}) and (\ref{18}), one
can then re-express solution VI as given by Eq. (\ref{2.29}) as
\bea\label{2.31}
&&u_1(x,t) = \frac{\beta}{\sqrt{2 g_{11}}}
[\sech(\beta x - \Delta) -\sech(\beta x + \Delta)]\,, \nonumber \\
&&u_2(x,t) = \frac{\beta}{\sqrt{2 g_{22}}}
[\sech(\beta x + \Delta)+ \sech(\beta x -\Delta)]\,,
\eea
where $B = \sinh^2(\Delta)$.

{\bf Superposed Solution VII}

It is easy to check that the coupled Eqs. (\ref{1}) and 
(\ref{2}) admit the superposed hyperbolic solution
\be\label{2.32}
u_1(x,t) = e^{i\omega_1 t} [1- \frac{A}{B+\cosh^2(\beta x)}]\,,
~~\phi_2(x) = e^{i\omega_2 t} \frac{D}
{B+\cosh^2(\beta x)}\,,~~A, B, D > 0\,,
\ee
provided
\bea\label{2.33}
&&g_{11} = -\omega_1 = 2\beta^2\,,~~A = \frac{3(1+2B)+\sqrt{8+(2B+1)^2}}{4}\,,
\nonumber \\
~&&g_{12} D^2 = [\sqrt{8+(2B+1)^2}-(2B+1)] \frac{3A\beta^2}{2}\,,
~~g_{21} + \omega_2 = 4\beta^2\,, \nonumber \\
&&g_{21} A = 3(2B+1) \beta^2\,,
~~g_{21} A^2 + g_{22} D^2 = 8B(B+1) \beta^2\,.
\eea
Thus for this solution $g_{11}, g_{12}, \omega_2, g_{21}$ are all positive, 
$\omega_1 < 0$ while the sign of $g_{22}$ depends on the value of $B$.  

On comparing the solution (\ref{2.32}) with the identity (\ref{12}), 
solution VII as given by Eq. (\ref{2.32}) can be
re-expressed as
\bea\label{2.34}
&&u_1(x,t) = e^{i\omega_1 t} [1- \frac{\beta}{\sqrt{2|g_{11}|}}
[\tanh(\beta x + \Delta)-\tanh(\beta x - \Delta)]]\,, \nonumber \\
&&u_2(x,t) = \frac{\sqrt{3}\beta}{2\sqrt{g_{12}}} K
[\tanh(\beta x + \Delta)-\tanh(\beta x - \Delta)]\,,
\eea
where $\sinh^2(\Delta) = B$ and 
\be\label{2.35}
K = \bigg [\cosh(2\Delta)\sqrt{8+\cosh^2(2\Delta)}
-[2+\cosh^2(2\Delta)] \bigg ]^{1/2}\,.
\ee

\section{Hyperbolic Superposed Solutions of the Coupled MKdV Model}

Let us consider the following coupled MKdV equations \cite{coupledmkdv} 
\be\label{3.1}
 u_{1t} + u_{1xxx} +6[g_{11} u_{1}^2 +g_{12} u_{2}^{2}] u_{1x} =0\,,
\ee
\be\label{3.2}
 u_{2t} + u_{2xxx} +6[g_{21} u_{1}^2 +g_{22} u_{2}^{2}] u_{2x} =0\,,
\ee

Before we discuss the superposed kink and pulse solutions, let us note
that the coupled Eqs. (\ref{3.1}) and (\ref{3.2}) admit the following kink
and pulse solutions. In particular, it is well known that Eqs. 
(\ref{3.1}) and (\ref{3.2}) admit a kink 
solution in both $u_1$ and $u_2$, i.e.
\be\label{3.3}
u_1(x,t) = A_1 \tanh(\xi)\,,
~~u_2(x,t) = A_2 \tanh(\xi)\,,~~\xi = \beta(x-vt)\,,
\ee
provided
\be\label{3.4}
v = -2\beta^2\,,~~g_{11} A_{1}^{2} 
+ g_{12} A_{2}^{2} = g_{21} A_{1}^{2} 
+ g_{22} A_{2}^{2} = -\beta^2\,.
\ee

It is also known that Eqs. (\ref{3.1}) and (\ref{3.2}) admit a pulse 
solution in both $u_1$ and $u_2$, i.e.
\be\label{3.5}
u_1(x,t) = A_1 \sech(\xi)\,,~~u_2(x,t) = A_2 \sech(\xi)\,,
\ee
provided
\be\label{3.6}
v = \beta^2\,,~~g_{11} A_{1}^{2} 
+ g_{12} A_{2}^{2} = g_{21} A_{1}^{2} 
+ g_{22} A_{2}^{2} = 2\beta^2\,.
\ee

Finally, Eqs. (\ref{3.1}) and (\ref{3.2}) are also known to admit 
a  mixed kink-pulse solution, i.e. say a kink solution in 
$u_1$ and a pulse solution in $u_2$ or vice versa. In 
particular, they admit
\be\label{3.7}
u_1(x,t) = A_1 \tanh(\xi)\,,~~u_2(x,t) = A_2 \sech(\xi)\,,
\ee
provided
\bea\label{3.8}
&&g_{12} A_{2}^{2} - g_{11} A_{1}^{2} 
= g_{22} A_{2}^{2} -g_{21} A_{1}^{2} = \beta^2\,, \nonumber \\ 
&&v= 6g_{21} A_{1}^2 + \beta^2 = 6 g_{12} A_{2}^{2} 
- 2 \beta^2\,.
\eea

We now show that the coupled Eqs. (\ref{3.1}) and (\ref{3.2}) 
admit four novel solutions which can be re-expressed
as the sum or the difference of the above kink solution.

We start with the ansatz
\be\label{3.9}
u_1(x,t) = \phi_1(\xi)\,,~~
u_2(x,t) = \phi_2(\xi)\,,~~\xi = \beta(x-vt)\,.
\ee
On substituting the ansatz (\ref{3.9}) in the 
coupled Eqs. (\ref{3.1}) and (\ref{3.2}) we get
\be\label{3.10}
\beta^2 \phi_{1\xi \xi \xi} = v \phi_{1\xi} 
-6[g_{11}\phi_{1}^{2}+g_{12}\phi_{2}^{2}]\phi_{1x}\,,
\ee
\be\label{3.11}
\beta^2 \phi_{2\xi \xi \xi} = v \phi_{2\xi} 
-6[g_{21}\phi_{1}^{2}+g_{22}\phi_{2}^{2}]\phi_{2x}\,.
\ee

We now show that the coupled Eqs. (\ref{3.10}) and (\ref{3.11}) 
and hence coupled Eqs. (\ref{3.1}) and (\ref{3.2}) admit four 
hyperbolic superposed solutions in terms of 
$\tanh(x+\Delta) \pm \tanh(x-\Delta)$. Out of these four solutions,
while the first solution is obtained when $u_2(x,t) \propto u_1(x,t)$, 
for the other three solutions, $u_1(x,t)$ and $u_2(x,t)$ are not 
proportional to each other. 

{\bf Solution I}

It is easy to check that 
\be\label{3.12}
u_1(x,t) = 1- \frac{A}{B+\cosh^2(\xi)}\,,~~u_2(x,t) 
= \alpha u_1(x,t)\,,~~A, B > 0\,,
\ee
is an exact solution to the coupled Eqs. (\ref{3.1}) and (\ref{3.2}), 
where $\alpha$ is any real number provided
\bea\label{3.13}
&&g_{11}+\alpha^2 g_{12} = g_{21}+g_{22} \alpha^2 
= -\frac{(2B+1)^2}{4B(B+1)} \beta^2\,,
\nonumber \\
&&A = \frac{4B(B+1)}{2B+1}\,,~~v= -\frac{(4B^2+4B+3)}
{2B(B+1)}\beta^2\,.
\eea
On comparing it with the identity (\ref{12}), we can re-express 
solution (\ref{3.12}) as
\be\label{3.14}
u_1(x,t)  = 1- \frac{\beta \tanh(2\Delta)}{\sqrt{|g_{11}+\alpha^2 g_{12}|}}
[\tanh(\xi+\Delta)-\tanh(\xi-\Delta)]\,,~~B = \sinh^2(\Delta)\,,
\ee
with $\xi = \beta(x-vt)$ and $u_2(x,t) = \alpha u_1(x,t)$ .

{\bf Solution II}

Yet another solution to the coupled Eqs. (\ref{3.1}) and (\ref{3.2})
is
\be\label{3.15}
u_1(x,t) = 1-\frac{A}{B+\cosh^2(\xi)}\,,~~u_2(x,t) = 
\frac{D\cosh(\xi) \sinh(\xi)}{B+\cosh^2(\xi)}\,,~~A,B,D > 0\,,
\ee
provided
\bea\label{3.16}
&&v = 4 \beta^2 +6 g_{11}+6g_{12} D^2\,,~~g_{11} +g_{12} D^2 
= g_{21}+ g_{22} D^2\,, \nonumber \\
&&g_{11} A^2 - B(2+B) g_{12} D^2 = 4B(B+1) \beta^2\,,~~
(2B+1) g_{11} A^2 + 2B(B+1) A g_{11}  \nonumber \\
&&= 2B(B+1)(2B+1) \beta^2\,,
g_{22}D^2 = -\frac{(10B^2+12B+3)}{(2B+1)^2} \beta^2\,, \nonumber \\
&&g_{21} A^2 = -\frac{B(B+1)(14B^2+12B+3)}{(2B+1)^2}\,,~~
A  = \frac{B(14B^2+12B+3)}{(2B+1)(B+1)} \beta^2\,.
\eea
Note that for this solution $g_{21}, g_{22}, g_{12} < 0$ while 
$g_{11} > 0$.

On comparing it with the novel identities (\ref{12}) and (\ref{13}) 
we can re-express solution (\ref{3.15}) as the superposed solution
\be\label{3.17}
u_1(x,t) = 1-\frac{\beta [\tanh(\xi+\Delta)-\tanh(\xi-\Delta)]}
{2\sinh(\Delta)\sqrt{|g_{21} \cosh(2\Delta)}}\,,
\ee
\be\label{3.18}
u_2(x,t) = \frac{\beta \sqrt{\cosh(\Delta)}}
{\sqrt{|g_{22}|}\cosh(2\delta)} 
[\tanh(\xi+\Delta)+\tanh(\xi-\Delta)]\,,
\ee
where $B = \sinh^2(\Delta)$.

{\bf Solution III}

Yet another hyperbolic superposed solution is 
\be\label{3.19}
\phi_1(\xi) = 1- \frac{A}{B+\cosh^2(\xi)}\,,~~\phi_2(\xi) 
= \frac{\alpha A}{B+\cosh^2(\xi)}\,, A, B > 0\,,
\ee
provided
\bea\label{3.20}
&&v = 4\beta^2 + 6 g_{11}\,,~~g_{21} = g_{11} < 0\,,~~g_{22} = g_{12}\,,
~~A g_{11} = -(2B+1)\beta^2\,, \nonumber \\
&&(g_{11}+\alpha^2 g_{12})A^2 = -4B(B+1) \beta^2\,.
\eea

On comparing it with the identity (\ref{12}), we can re-express 
solution (\ref{3.19}) as
\be\label{3.21}
u_1(x,t) = 1- \frac{\beta}{\sqrt{|g_{11}+\alpha^2 g_{22}|}}
[\tanh(\xi+\Delta) -\tanh(\xi-\Delta)]\,,
\ee
and 
\be\label{3.22}
u_2(x,t)  = \frac{\alpha \beta}{\sqrt{|g_{11}+\alpha^2 g_{22}|}}
[\tanh(\xi+\Delta) -\tanh(\xi-\Delta)]\,,~~B = \sinh^2(\Delta)\,,
\ee
where $B = \sinh^2(\Delta)$.

{\bf Solution IV}

Yet another hyperbolic superposed solution to the coupled Eqs. 
(\ref{3.1}) and (\ref{3.2}) is 
\be\label{3.23}
u_1(x,t) = 1- \frac{A}{B+\cosh^2(\xi)}\,,~~\phi_2(\xi) 
= -b -\frac{\alpha A}{B+\cosh^2(\xi)}\,,~~A, B, b > 0
\ee
provided
\bea\label{3.24}
&&v-4\beta^2 = g_{12} + b^2 g_{22} = g_{11}+b^2 g_{12}\,,
\nonumber \\
&&A(g_{11} - g_{12} \alpha b) = = A(g_{21}-g_{22} \alpha b)
= -(2B+1)\beta^2\,, \nonumber \\
&&(g_{11}+\alpha^2 g_{12}) A^2 = 
= (g_{21}+\alpha^2 g_{22}) = -4B(B+1) \beta^2\,. 
\eea

On comparing it with the novel identity, we can re-express 
solution (\ref{3.22}) as
\be\label{3.25}
u_1(x,t)  = 1- \frac{\beta}{\sqrt{|g_{11}+\alpha^2 g_{22}|}}
[\tanh(\xi+\Delta) -\tanh(\xi-\Delta)]\,,
\ee
and 
\be\label{3.26}
u_2(x,t) = -b - \frac{\alpha \beta}{\sqrt{|g_{11}+\alpha^2 g_{22}|}}
[\tanh(\xi+\Delta) -\tanh(\xi-\Delta)]\,,
\ee
where $B = \sinh^2(\Delta)$.

\section{Conclusion and Open Problems}

In this paper we have considered a coupled $\phi^4$ \cite{coupledphi4}, a coupled 
NLS \cite{manakov} and a coupled MKdV \cite{coupledmkdv} model and demonstrated 
that all of them not only admit single kink and single pulse solutions but also admit novel 
solutions which can be re-expressed in terms of the sum or the difference of two kink 
or two pulse solutions. For the coupled $\phi^4$ and the coupled NLS models, we have 
obtained six superposed solutions essentially covering all possible cases among 
$\tanh(x+\Delta) \pm \tanh(x-\Delta)$ and $\sech(x-\Delta) \pm \sech(x+\Delta)$ 
with the two coupled members (or fields) being distinct. On the other hand, for the coupled 
MKdV we have only one such solution involving $\tanh(x+\Delta) \pm \tanh(x-\Delta)$.
Besides, we have obtained one superposed solution each of the coupled $\phi^4$ 
and the coupled NLS models and three superposed solutions of the coupled MKdV 
model where both the coupled members (or fields) involve only 
$\tanh(x+\Delta)-\tanh(x-\Delta)$.

This paper raises several questions some of which are:  

\begin{enumerate}

\item How many of these superposed solutions are stable in each coupled model? 
If stable, where can one look for such structures experimentally, e.g. in photonics 
\cite{mecozzi}? What is the connection of such superposed solutions vis a vis a
single kink or a single pulse solution?

\item So far, except for $\tanh(x+\Delta) - \tanh(x-\Delta)$, no 
one has been able to find superposed hyperbolic solutions of the form
$\tanh(x+\Delta)+\tanh(x-\Delta)$ or of the form $\sech(x+\Delta) \pm
\sech(x-\Delta)$ in any {\it uncoupled} model. It would be worthwhile 
finding such solutions in uncoupled models. This will enable the physical 
interpretation of such superposed solutions in comparison to the
corresponding single kink or single pulse solution.

\item Presumably there are other coupled systems, e.g. coupled KdV arising in two-layer 
fluids \cite{ckdv}, which might admit solutions similar to those presented in this paper. 
It would be desirable to find such models and the corresponding  superposed solutions. 

\item The three coupled models considered in this paper will obviously also admit 
superposed periodic kink and pulse solutions in terms of Jacobi elliptic functions. It 
would be interesting to find such solutions and see how many of them smoothly go 
over to the hyperbolic superposed solutions obtained in this paper. We hope to address 
this issue in the near future.

\end{enumerate}

\noindent{\bf Acknowledgment}

One of us (AK) is grateful to Indian National Science Academy (INSA) for the
award of INSA Senior Scientist position at Savitribai Phule Pune University. 
The work at Los Alamos National Laboratory was carried out under the auspices of 
the U.S. DOE and NNSA under Contract No. DEAC52-06NA25396.


\begin{thebibliography}{99}

\bibitem{ks22} A. Khare and A. Saxena, arXiv: 2202.06223 (2022).

\bibitem{tan} A.P. Tanakeyav, V.V. Smagin, M.A. Borich and A.S. Zhuravlev, 
The Physics of Metals and Metallography {\bf 107} (2009) 229; ibid. {\bf 107} (2009) 245.   

\bibitem{sma} V.V. Smagin, A.P. Tanakeyav and M.A. Borich, The Physics of 
Metals and Metallography {\bf 108} (2009) 425.   

\bibitem{dashen} R. F. Dashen, B. Hasslacher and A. Neveu, 
Phys. Rev. D {\bf 12} (1975) 2443. 

\bibitem{campbell} D. K. Campbell and A. R. Bishop, Nucl. Phys. B {\bf 200} 
(1982) 297. 

\bibitem{saxena} A. Saxena and A. R. Bishop, Phys. Rev. A {\bf 44} (1991) R2251. 

\bibitem{thies} O. Schnetz, M. Thies and K. Urlichs, Annals Phys. {\bf 321} 
(2006) 2604. 

\bibitem{abel} T. Abel and R. Siems, Ferroelectrics {\bf 153} (1994) 177. 

\bibitem{curnoe} S.H. Curnoe and I. Munawar, Physica B {\bf 378-380} (2006) 554. 

\bibitem{mecozzi} A. Mecozzi, C. Antonelli and M. Shtaif, Opt. Exp. {]bf 20} (2012) 23436. 

\bibitem{coupledphi4} A. Khare and A. Saxena, J. Math. Phys. {\bf 47} (2006) 092902. 

\bibitem{manakov} S. V. Manakov, Sov. Phys. JETP {\bf 38} (1973) 248; 
V. E. Zhakharov and S. V. Manakov, Sov. Phys. JETP {\bf 42} (1976) 842.  

\bibitem{coupledmkdv}  E. Fan, Phys. Lett. A {\bf 282} (2001) 18. 

\bibitem{ckdv} S. Y. Lou, B. Tong, H.-C. Hu and X.-Y. Tang, J. Phys. A {\bf 39} (2006) 513. 

\end{thebibliography}
\end{document}